\documentclass[prl,twocolumn,aps,nofootinbib,superscriptaddress]{revtex4}
\usepackage{amssymb}
\usepackage{amsmath}
\usepackage{amsfonts}
\usepackage{graphicx}
\usepackage{bm}
\usepackage{epsfig}

\newif\ifpdf
\ifx\pdfoutput\undefined
\pdffalse 
\else
\pdfoutput=1 
\pdftrue
\fi

\def\Dslash{D\!\!\!\!\slash}

\def\bnslash{\bar n\!\!\!\slash}

\def\OMIT#1{}

\newcommand{\nn}{\nonumber}

\newcommand{\bn}{{\bar n}}
\newcommand{\bea}{\begin{eqnarray}}
\newcommand{\eea}{\end{eqnarray}}

\newcommand{\bnP}{\bar {\cal P}}

\newcommand{\mcdot}{\!\cdot\!}

\begin{document}

\preprint{\vbox{\hbox{MIT-CTP 3445}  \hbox{hep-ph/0311285}
}}
\title{ Model Independent Results for SU(3) Violation\\
in Light-Cone Distribution Functions }
\author{Jiunn-Wei Chen}
\affiliation{Department of Physics, National Taiwan University, Taipei 10617, 
Taiwan 
\thanks{
Electronic address: jwc@phys.ntu.edu.tw}}
\author{Iain W. Stewart\vspace{0.3cm}}
\affiliation{CTP, Massachusetts Institute for Technology,
Cambridge, MA 02139, USA\thanks{
Electronic address: iains@mit.edu} \vspace{0.2cm} }

\begin{abstract}
  
  Using chiral symmetry we investigate the leading SU(3) violation in light-cone
  distribution functions $\phi _{M}(x)$ of the pion, kaon, and eta. It is shown
  that terms non-analytic in the quark masses do not affect the shape, and only
  appear in the decay constants.  Predictive power is retained including the
  leading analytic $m_{q}$ operators.  With the symmetry violating corrections
  we derive useful model independent relations between $\phi _{\pi}$, $\phi
  _{\eta }$, $\phi _{K^{+},K^{0}}$, and $ \phi _{\bar{K}^{0},K^{-}}$.  Using the
  soft-collinear effective theory we discuss how factorization generates the
  subleading chiral coefficients.

\end{abstract}

\maketitle

\ifpdf
\DeclareGraphicsExtensions{.pdf, .jpg} \else
\DeclareGraphicsExtensions{.eps, .jpg} \fi

At large momentum transfers $Q^{2}\gg \Lambda _{\mathrm{QCD}}^{2}$,
factorization in QCD dramatically simplifies hadronic processes. Observables
depend on universal non-perturbative light cone distribution functions like
$\phi _{M}(x,\mu )$ for exclusive processes with mesons $M$, or the parton
distribution functions $f_{i/H}(x,\mu )$ for deep inelastic scattering on hadron
$H$. A well known example is the electromagnetic form factor~\cite{bl,cz}, which
dropping $\Lambda_{\rm QCD}/Q$ corrections is
\begin{eqnarray}  \label{F1}
  F(Q^2) = \frac{f_a f_{b}}{Q^2}\! \int_0^1 \!\!\!\! dx\, dy\: T(x,y,Q,\mu)
  \phi_a(x,\mu) \phi_{b}(y,\mu) \,.
\end{eqnarray}
Here $f_{i}$ are meson decay constants and the hard scattering kernel $T$ is
calculated perturbatively. For $a=b=\pi ^{\pm }$, $T(x,y,\mu =Q)=8\pi \alpha
_{s}(Q)/(9xy)+\mathcal{O}(\alpha _{s}^{2})$, which leads to inverse moments of
the light-cone distribution functions. The same distributions $\phi_{a}$, appear
in many factorization theorems including those relevant to measuring fundamental
parameters of the Standard Model~\cite{Bprocesses}, such as $B\rightarrow \pi
\ell \nu, \eta\ell\nu $ which give the CKM matrix element $|V_{ub}|$,
$B\rightarrow D\pi $ used for tagging, and $B\rightarrow \pi \pi ,K\pi
,K\bar{K}, \pi\eta, \ldots$ which are important for measuring CP violation.

QCD factorization theorems like the one in Eq.~(\ref{F1}) make no attempt to
separate the light quark mass scales $m_{u,d,s}$ from scales such as $ \Lambda
_{\mathrm{QCD}}$ or $\Lambda_{\mathrm{\chi }}$ (the scale of chiral symmetry
breaking). Thus $\phi _{a}(x,\mu )$ is a function of all of these scales. In the
chiral limit $m_{u,d,s}\rightarrow 0$, SU(3)  symmetry predicts a further
relation $\phi _{\pi }=\phi _{K}=\phi _{\eta }=\phi _{0}$. For simplicity we
work in the isospin limit and the $\overline{\mathrm{MS}} $ scheme, and
normalize the distributions so that $\int \!\!dx\,\phi_{M}(x)\!=\!1$.
Generically from chiral symmetry the leading order SU(3) violation takes the
form $[M\!=\!\pi ,K,\eta ]$
\begin{eqnarray}
  \phi_{M}(x,\mu)\!\!&=&\!\!\phi_{0}(x,\mu) 
  +\!\!\!\!\! \sum\limits_{P=\pi ,K,\eta }\!\frac{m_{P}^{2}}{(4\pi f)^{2}}
  \Big[ E_{M}^{P}(x,\mu)\ln \Big(\frac{m_{P}^{2}}{\mu_{\!\chi}^{2}}\Big) 
  \notag \\ &&
  +F_{M}^{P}(x,\mu,\mu_{\!\chi} )\Big]\ .
\end{eqnarray} 
The functions $\phi_{0}$ , $E_{M}^{P}$, and $F_{M}^{P}$ are independent of
$m_{q}$, and are only functions of $\Lambda_{\mathrm{QCD}}$, $\mu $, and $x$ the
momentum fraction of the quark in the meson at the point of the hard
interaction. $F_{M}^{P}$ also depends on the ChPT dimensional regularization
parameter $\mu_{\!\chi}$ which cancels the $\ln(m_P^2/\mu_{\!\chi}^2)$
dependence, so by construction ${\phi_{M}}$ is $\mu_{\!\chi}$ independent.  So
far these sizeable $\sim 30\%$ corrections to the SU(3) limit have mostly been
estimated in a model dependent fashion and even the signs of the effects remain
uncertain. Recent light-cone sum rule results give the ratio of moments $\langle
x_{u}^{-1}\rangle_{\pi ^{+}}/\langle x_{u}^{-1}\rangle_{K^{+}}\simeq
1.12$~\cite{ball}, correcting the earlier result $\langle x_{u}^{-1}\rangle_{\pi
  ^{+}}/\langle x_{u}^{-1}\rangle_{K^{+}}\simeq 0.80$~\cite{cz,ball2}. In $B\to
MM'$ decays SU(3) violation was studied in~\cite{Bsu3}.

In this letter we prove, by using chiral perturbation theory (ChPT), that at
first order the light-cone distributions are analytic in $m_{q}$, meaning that
\begin{equation}
  E_{M}^{\pi }(x)=0\,,\qquad E_{M}^{K}(x)=0\,,\qquad E_{M}^{\eta }(x)=0\,.
\end{equation} 
Thus, chiral logarithms appear only in the decay constants $f_{M}$. We also
derive relations between moments of the $F_{M}^P(x)$ coefficients, and thus
determine model independent results for $\phi_{M}(x)$ that are valid at first
order in the SU(3) violation, i.e. at the 10\% level. Finally, we discuss quark mass
effects using factorization and the soft-collinear effective theory~\cite{SCET}
(SCET), and explain how ChPT and SCET results can be combined.

Recently ChPT has been applied to the computation of hadronic twist-2 matrix
elements~\cite{AS,CJ}.  Many applications have been worked out, e.g.~chiral
extrapolations of lattice data, generalized parton distributions~\cite{Jq},
large $N_{C}$ relations among distributions, and soft pion productions in deeply
virtual Compton scattering~\cite{DVCSpi}.  It appears likely to us that parton
SU(3) violation will be first quantitatively measured in meson light-cone
distributions. This letter provides the necessary framework.


In momentum space the light-cone distribution functions can be defined by $ 
\langle M^{b}|O^{A,a}(\omega _{+},\omega _{-})|0\rangle $, via
\begin{eqnarray}\label{phidefn}
  &&\langle M^{b}|(\bar{\psi}Y)_{\omega _{1}}n\!\!\!\slash\gamma _{5}
  \lambda^{a}(Y^{\dagger }\psi )_{\omega _{2}}|0\rangle =-i\delta ^{ab}
  \delta \Big( \frac{\omega _{-}\!-\!n\!\cdot \!p_{M}}{2}\Big)  \notag   \\
&& \times f_{M}\,n\!\cdot \!p_{M}\!\int_{0}^{1}\!\!\!\!dx\,\delta 
  (\omega_{+}\!-\!(2x\!-\!1)\,n\!\cdot \!p_{M})\phi _{M}(x,\mu )\,,
\end{eqnarray} 
where $n$ is a light-like vector, $n^{2}=0$, $\omega _{\pm }=\omega _{1}\!\pm
\!\omega _{2}$, and our octet matrices are normalized so that $
\mathrm{tr}[\lambda ^{a}\lambda ^{b}]=\delta ^{ab}$. We use the subscript
notation from SCET, so that $(\bar{\psi}Y)_{\omega _{1}}$ is the Fourier
transform of $\bar{\psi}(x^{-})Y(x^{-},\infty )$, where $Y$ is a Wilson line of
$n\!\cdot \!A$ gluons. Therefore $\omega _{1}$ is the $n\!\cdot \!p$ momentum
carried by this gauge invariant product of fields. Hard perturbative corrections
will generate convolutions with coefficients $C(\omega _{\pm },Q^{2},\mu )$.  \OMIT{At
  the scale $\mu =\omega_{+}$} We can expand $C(\omega
_{+},Q^{2})=\sum_{k=0}^{\infty }C_{k}(-\omega _{+})^{k}$ so that
\begin{eqnarray}
  \int\!\! d\omega _{+}\: C(\omega _{+})\, O^{A,a}(\omega_{+})
   =\sum_{k=0}^{\infty }C_{k}\, O_{k}^{A,a}\,,
\end{eqnarray} 
where the tower of octet axial twist-2 operators are 
\begin{eqnarray}
  O_{k}^{A,a}=\overline{\psi }\, n\!\!\!\slash\gamma _{5}\lambda ^{a}\big[ 
  in\!\cdot \!\overleftrightarrow{D}\,\big]^{k}\,\psi \,.  \label{phimoments}
\end{eqnarray} 
Here $i\overleftrightarrow{D}=i\overleftarrow{D}-i\overrightarrow{D}$ and
having the vector indices dotted into $n^{\mu _{1}}\cdots n^{\mu _{k+1}}$ has
automatically projected onto the symmetric and traceless part. Comparing
Eqs.~(\ref{phidefn}) and (\ref{phimoments}) gives [$z=1-2x$] 
\begin{eqnarray}  \label{me} 
&& \langle M^{b}|O_{k}^{A,a}|0\rangle 
  = -if_{M}\delta ^{ab}(n\!\cdot\!p_{M})^{k+1}\big\langle z^{k}\big\rangle_{M}\,,
    \notag \\
&& \big\langle z^{k}\big\rangle_{M}
  = \int_{0}^{1}\!\!\!dx\,\big(1-2x\big)^{k}\,\phi _{M}(x)\,.
\end{eqnarray} 
Thus, the matrix element of $O_{k}^{A,a}$ is related to moments of the meson
light-cone distribution functions.  A subscript $M=\eta$ denotes the purely
octet part.  It is convenient to work with $O_{k}^{A,a}=O_{k}^{R,a}-O_{k}^{L,a}$
where
$
  O^{R,a}_k =  
  \overline{\psi }_{R}n\!\!\!\slash\lambda _{R}^{a}\big[in\!\cdot \!
   \overleftrightarrow{D}\big]^{k}\psi _{R}\,,  
  O^{L,a}_k =
    \overline{\psi }_{L}n\!\!\!\slash\lambda _{L}^{a}\big[in\!\cdot \! 
    \overleftrightarrow{D}\big]^{k}\psi _{L}\,,
$
and $\psi _{L,R}=(1\mp \gamma_{5})/2\,\psi $. The distinction between
$\lambda_{R}^{a}$, $\lambda_{L}^{a}$, and $\lambda ^{a}$ is for bookkeeping
purposes, and we set $\lambda_{R,L}^{a}=$ $\lambda ^{a}$ at the end.

When $a=3$ or $8$, $O_{k}^{A,a}$ transforms simply under charge conjugation
($\mathcal{C}$), being even when $k$ is even, and odd when $k$ is odd. The meson
states $\pi ^{0}$ and $\eta $ (ie. $M^{3,8}$) are $\mathcal{C}$ even.  Thus from
Eq.~(\ref{me}), $\left\langle z^{k}\right\rangle_{\pi ^{0},\eta }$ vanish when
$k$ is odd due to $\mathcal{C}$ (and using isospin the same applies for
$M=\pi^{\pm}$). For all $a$'s the operator would transform as
\begin{eqnarray}
   \mathcal{C}^{-1}O_{k}^{A,a}\mathcal{C}=(-1)^{k}O_{k}^{A,a}\,,
\end{eqnarray} 
if we demanded that under the $\mathcal{C}$ transformation 
\begin{eqnarray}
 \lambda _{R}^{a}\rightarrow \lambda _{L}^{a\,T}\,,\qquad 
   \lambda_{L}^{a}\rightarrow \lambda _{R}^{a\,T}\,.  \label{lambdac}
\end{eqnarray} 
Eq.~(\ref{lambdac}) will be used to reproduce the $\mathcal{C}$ violating
properties of $O_{k}^{A,a}$ when matching to the hadronic operators.


To construct the hadronic ChPT operators we define $\Sigma =\exp (2i\pi
^{a}\lambda ^{a}/f)$ and $m_{q}=\mathrm{diag}(\overline{m},\overline{m}
,m_{s})=m_{q}^{\dagger }$. Under a chiral $SU(3)_{L}\!\times
\!SU(3)_{R}$ transformation
\begin{eqnarray}
 \Sigma &\rightarrow &L\Sigma R^{\dagger },\qquad \quad m_{q}\rightarrow
   L m_{q} R^{\dagger }\,,  \notag \\
\lambda _{R}^{a} &\rightarrow & R\lambda _{R}^{a} R^{\dagger }, \qquad 
   \lambda_{L}^{a}\rightarrow L\lambda _{L}^{a}L^{\dagger }.
\end{eqnarray} 
Under charge conjugation $\Sigma \rightarrow \Sigma ^{T}$, while
$\lambda_{R,L}^{a}$ transform according to Eq.~(\ref{lambdac}). At next to
leading order (NLO) in the $p^2/\Lambda_\chi^2$ and $m_M^2/\Lambda_\chi^2$
chiral expansion
\begin{equation}
  O_{k}^{A,a}\longrightarrow \ \ \sum_{i}c_{k,i}\mathcal{O}_{k,i}^{A,a}+ 
  \sum_{i}b_{k,i}\overline{\mathcal{O}}_{k,i}^{A,a}\,,
\end{equation} 
where the $\mathcal{O}$'s are leading order (LO) and the
$\overline{\mathcal{O}}$'s are NLO. The sum on $i$ runs over hadronic
operators having the same transformation properties as $O_{k}^{A,a}$. The ChPT
Wilson coefficients $c_{k,i}$ and $b_{k,i}$ encode physics at the scale
$p^{2}\sim \Lambda_{\chi }^{2}$ and the operators encode $p^{2}\ll \Lambda_{\chi
}^{2} $.


At LO in the chiral expansion only one operator contributes in our analysis
\begin{eqnarray}
\mathcal{O}_{j-1,0}^{A,a} &=&
  \frac{f^{2}}{8}\,\mathrm{Tr}\Big[\lambda_{R}^{a}\Big\{\Sigma ^{\dagger }
  \Box ^{j}\Sigma +(-1)^{j}(\Box ^{j}\Sigma^{\dagger })\Sigma \Big\}  \notag \\
&& \hspace{-0.5cm}
  -\lambda _{L}^{a}\Big\{\Sigma \Box ^{j}\Sigma ^{\dagger}+(-1)^{j}
  (\Box ^{j}\Sigma )\Sigma ^{\dagger }\Big\}\Big]\,,
\end{eqnarray}
where $\Box^{j}=(in\cdot \partial )^{j}$ and the factors of $f$ have been
inserted by using chiral counting rules ($f=f_M$ in the chiral limit). All other
$\mathcal{O} _{k,i}^{A,a}$ operators have $\Box ^{k+1}$ factors acting on more
than one $ \Sigma $ field and as we will see, do not contribute to the
$0\rightarrow M^{b}$ matrix element up to NLO.

Under charge conjugation, $\mathcal{O}_{k,0}^{A,a}\rightarrow (-1)^{k}
\mathcal{\ O}_{k,0}^{A,a}$. Thus, when $k$ is odd this operator gives vanishing
$0\rightarrow M^{b}$ matrix elements, as expected by $\mathcal{C}$ and SU(3)
symmetry. For $k$ even
\begin{equation}
  \langle M^{b}|c_{k,0}\mathcal{O}_{k,0}^{A,a}|0\rangle 
  =-if_{M}\delta^{ab}(n\!\cdot \!p_{M})^{k+1}c_{k,0}\,,
\end{equation}
and comparing with Eq.~(\ref{me}) we see that 
\begin{equation}
  c_{k,0} = \langle z^{k}\rangle _{0}
   =\int_{0}^{1}\!\!\!\!dx\,(1\!-\!2x)^{k} \phi_{0}(x)\,,
\end{equation}
where $\phi_{0}$ is the distribution function in the SU(3) limit. Note that
$c_{0,0}=1$ due to our normalization for $\phi_M$.

\begin{figure}[t!]
\vskip 0.cm 
\centerline{
  \mbox{\epsfxsize=5.6truecm \hbox{\epsfbox{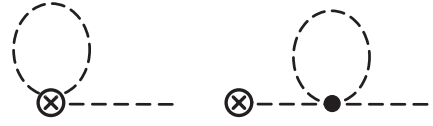}} }
  } \vskip-0.1cm
\caption[1]{NLO loop diagrams, where here $\otimes$ denotes an insertion of
  $\mathcal{O}_{k,0}^{A,a}$, and the dashed lines are meson fields.}
\label{fig:loops}\vspace{-0.5cm}
\end{figure}
At NLO chiral logarithms can be obtained from loop diagrams involving the LO
operators as shown in Fig.\,\ref{fig:loops}. For $k=0$ the operator $
O_{k=0}^{A,a}$ is the axial current, while $\mathcal{O}_{k=0,0}^{A,a}$ is just
the standard ChPT axial current operator whose Fig.\,\ref{fig:loops} graphs give
the one-loop corrections to $f_M$. For odd $k$ the one-loop graphs vanish since
adding the chiral Lagrangian does not change the $\mathcal{C}$-invariance
argument. For any $k>0$ the diagrams have a term where all derivatives act on
the outgoing meson line, and this gives the same corrections as for $f_{M}$.
The first diagram could have additional contributions from derivatives acting
inside the loop but it is straightforward to show that these diagrams vanish
identically since $n^2=0$, and that the same holds true for LO operators with
derivatives on more than one $\Sigma $~\cite{AS}. Thus, we have shown that all
possible non-analytic corrections are contained in $f_{M}$ at NLO. This is true
for every moment, and so we conclude that the leading order SU(3) violation of
$\phi_{M}(x)$ is analytic in ${m_{q}}$.


\begin{widetext}
Analytic corrections are also generated by subleading operators, and at NLO we
find the basis [$B_0=-2 \langle \bar\psi \psi \rangle/f^2$]
\begin{eqnarray} \label{NLO}
 \overline {\cal O}_{j-1,1}^{A,a} &=& 
  2 B_0\: {\rm Tr}\big[m_{q}\Sigma ^{\dagger }+\Sigma m_{q}^{\dagger }\big]\
  {\rm Tr} \Big[ \lambda _{R}^{a} \Big\{
    \Sigma^{\dagger } \Box^{j} \Sigma  + (-1)^{j}( \Box^{j} \Sigma^{\dagger })
   \Sigma \Big\}  
   -\lambda _{L}^{a}\Big\{  \Sigma \Box^{j} \Sigma^{\dagger }  
    + (-1)^{j} ( \Box^{j} \Sigma ) \Sigma^{\dagger }\Big\} \Big] 
  \,, \nn \\
 \overline {\cal O}_{j-1,2}^{A,a} &=&
   2 B_0\: {\rm Tr}\Big[ 
   \lambda_{R}^{a}\big\{  m_{q}^{\dagger} \Box^j \Sigma  
     + (-1)^{j} \Box^j \Sigma ^{\dagger } m_{q} \big\} 
   - \lambda_{L}^{a}\big\{  m_{q} \Box^j \Sigma^\dagger  
     + (-1)^{j} \Box^j \Sigma  m_{q}^\dagger  \big\} 
   \big] \,.
\end{eqnarray} 
\end{widetext}
All other NLO operators have derivatives on more than one $\Sigma$, or can be
reduced to $\overline{\mathcal{O}}_{j-1,1}^{A,a}$ and
$\overline{\mathcal{O}}_{j-1,2}^{A,a}$ using the equations of motion. For
instance, consider
\begin{eqnarray}
  \overline{\mathcal{O}}_{j-1,3}^{A,a} \!\!&=&\!\! 
   2 B_0 \mathrm{Tr}\Big[\lambda _{R}^{a}  
   \big\{ \Sigma ^{\dagger }m_{q}\Sigma ^{\dagger }\Box ^{j}\Sigma
   \!+\! (-1)^{j} (\Box ^{j}\Sigma ^{\dagger })\Sigma m_{q}^{\dagger }\Sigma \big\}  
   \notag \\
&& \hspace{-1cm}-\lambda _{L}^{a}\big\{ \Sigma m_{q}^{\dagger }\Sigma
  \Box ^{j}\Sigma ^{\dagger }\!+\! (-1)^{j} (\Box ^{j}\Sigma )\Sigma ^{\dagger }m_{q}
  \Sigma^{\dagger }\big\}\Big]\,.
\end{eqnarray}
The sum and difference $\overline{\mathcal{O}}_{k,2}^{A,a}\pm
\overline{\mathcal{O}}_{k,3}^{A,a}$ contain factors of $(\Sigma^{\dagger} m_{q}
\pm m_{q}^{\dagger}\Sigma )$ and $(\Sigma m_{q}^{\dagger}\pm m_{q}
\Sigma^{\dagger})$. Using the equations of motion for $\Sigma$
\begin{equation}
  \Sigma ^{\dagger }(i\partial _{\mu })^{2}\Sigma 
   =-(i\partial ^{\mu }\Sigma^{\dagger })
   (i\partial_{\mu}\Sigma )\!+\! B_0 (\Sigma^{\dagger}m_{q}\!
   -\!m_{q}^{\dagger }\Sigma )\,,  \label{eom}
\end{equation}
together with the analogous equation for $\Sigma^{\dagger }$ we can trade
$\overline{\mathcal{O}}_{k,2}^{A,a}-\overline{\mathcal{O}}_{k,3}^{A,a}$ for
operators with derivatives on more than one $\Sigma $. These additional
operators do not generate one-meson matrix elements at tree level and can be
omitted from our analysis. Thus only $\overline{\mathcal{O}}_{k,2}^{A,a}+
\overline{\mathcal{O}}_{k,3}^{A,a}$ contributes and for simplicity we trade this
for $\overline{\mathcal{O}}_{k,2}^{A,a}$. We can also consider operators
analogous to $\overline{\mathcal{O}}_{k,3}^{A,a}$ but with the $ \Box^{k}$
factors on a different $\Sigma $. Since $\Box ^{k}(\Sigma^{\dagger }\Sigma )=0$
we can use
\begin{equation}
  0 = (\Box ^{k}\Sigma ^{\dagger })\Sigma +\Sigma ^{\dagger }(\Box ^{k}\Sigma )
    +\ldots \,,  \label{moveBox}
\end{equation}
where the ellipse denote $(\Box ^{k-m}\Sigma^{\dagger })(\Box ^{m}\Sigma )$
terms that only contribute for matrix elements with more than one meson.  Thus,
Eq.~(\ref{moveBox}) allows us to move factors of $\Box ^{k}$ onto a neighboring
$\Sigma $ and eliminate operators. Finally, we can consider operators where the
power suppression is generated by derivatives rather than a factor of $m_{q}$.
Boost invariance requires that these operators still have $j$ factors of $n^{\mu
}$, so they will involve $\Box^{j}$ just like the operators we have been
considering. To get power suppression with derivatives we can either use
$(\partial_{\mu }\Sigma^{\dagger })(\partial^{\mu }\Sigma )$ which has
derivatives on more than one $\Sigma $, or $ \Sigma^{\dagger }(\partial^{\mu
})^{2}\Sigma $ which can be traded for operators with $m_{q}$'s using
Eq.~(\ref{eom}).  Therefore, the operators with $m_{q}$'s in Eq.~(\ref{NLO})
suffice.


At NLO we need $\langle \overline{\mathcal{O}}_{k,i}^{A,a}\rangle$ at tree
level.  The $k=0$ operators are equivalent to those derived from the standard ${\mathcal
  O}(p^{4})$ chiral Lagrangian\OMIT{~\cite{GL}}, so $b_{0,1}=L_{4}$ and
$b_{0,2}=L_{5}$. Adding the wavefunction counterterms we find $\langle \pi
^{b}|c_{k,0} Z^{1/2} \mathcal{O}_{k,0}^{A,a} + \sum_{i} b_{k,i}
\overline{\mathcal{ O}}_{k,i}^{A,a} | 0 \rangle = N_{k} A_{k}$, where
$N_{k}=-if(n\!\cdot \!p_{M})^{k+1}$ and the NLO contribution is
\begin{eqnarray} \label{Ak0} 
  A_{k}\!\! &=&\!\!\frac{4B_0}{f^{2}} \Big\{{\big[1\!-\!(-1)^{k+1}\big]}2\mathrm{Tr} 
  [m_{q}]\delta ^{ab}(2b_{k,1}\!-\!L_{4}c_{k,0})\phantom{xx} \notag \\
&&\hspace{-0.9cm}+\mathrm{Tr}\big[m_{q}\{\lambda^{a}
  \lambda^{b}\!-\!(-1)^{k+1}\lambda ^{b}\lambda ^{a}\}\big](2b_{k,2}\!
  -\!L_{5}c_{k,0})  \Big\}\,.  
\end{eqnarray} 
For $k=0$, $A_0$ gives the standard counterterm corrections to $f_{M}$ at NLO,
which are combined with the one-loop contributions from Fig.~\ref{fig:loops}.
Numerically
\vspace{-0.2cm}
\begin{equation}
\vspace{-0.2cm}
\frac{f_{K}}{f_{\pi }}=1.23\,,\qquad \frac{f_{\eta }}{f_{\pi }}=1.33\,.
\end{equation}
 To compute the analytic
chiral corrections to the moments of $\phi _{M}(x)$ we subtract the corrections
to $f_{M}$,
\begin{eqnarray} \label{Ak} 
\Delta A_{k}\!\! &=&\!\! \frac{8B_0}{f^{2}}  \Big\{{\big[1\!-\!(-1)^{k+1}\big]}
2\mathrm{Tr}[m_{q}]\delta ^{ab}(b_{k,1}\!-\!L_{4}c_{k,0})\phantom{xx}
\notag \\
&&\hspace{-0.9cm}+\mathrm{Tr}\big[m_{q}\{\lambda ^{a}\lambda
^{b}\!-\!(-1)^{k+1}\lambda ^{b}\lambda ^{a}\}\big](b_{k,2}\!-\!L_{5}c_{k,0}) 
\Big\}\,. 
\end{eqnarray}

For odd $k$ (odd moments), $\Delta A_k$ collapses to the $\mathrm{Tr} \big[
m_{q} \big[ \lambda^{a}, \lambda^{b} \big] \big]$. This yields [$m=0,1,2,\cdots $]
\begin{eqnarray} \label{oddMM}
\vspace{-0.3cm}
 \big\langle z^{2m+1}\big\rangle_{\pi }
   \!&=&\! \big\langle z^{2m+1}\big\rangle_{\eta}=0\ , \\
  \big\langle z^{2m+1}\big\rangle_{K^{0}}
  \!\! &=&\!\! \big\langle z^{2m+1}\big\rangle_{K^{+}}
  \!=\! -\big\langle z^{2m+1} \big\rangle_{\overline{K}^{0}} 
   \!=\! -\big\langle z^{2m+1}\big\rangle_{K^{-}}  \notag \\
  && \hspace{-1cm}   
   = \frac{8 B_0 (m_s\!-\! \bar m)}{f^2}  \  
   b_{2m+1,2}
  \,. \nn
\end{eqnarray}
The leading SU(3) violation for odd $k$ agrees with our expectations based on
${\mathcal C}$ and isospin symmetry.  For even $k$ (even moments), the $\Delta
A_{k}$ gives structures $\delta ^{ab}\mathrm{Tr}[m_{q}]$ and
$\mathrm{Tr}[m_{q}\{\lambda ^{a},\lambda ^{b}\}] $ so
\begin{eqnarray} \label{evenMM}
 \delta \big\langle z^{2m}\big\rangle_{\pi } 
  &=& 2\overline{m}\, \alpha _{2m}+(2 \overline{m}+m_{s})\beta _{2m}\ , 
  \\
\delta \big\langle z^{2m}\big\rangle_{K} 
  &=& (\overline{m}+m_{s})\alpha_{2m}+(2\overline{m}+m_{s})\beta _{2m}\ ,
   \label{even} \nn \\
\delta \big\langle z^{2m}\big\rangle_{\eta } 
  &=& \frac{(2\overline{m}+4m_{s})}{3}\, \alpha _{2m}+(2\overline{m}+m_{s})
   \beta_{2m}\ ,  \notag
\end{eqnarray}
where $\delta$ means the deviation from the chiral limit value, \OMIT{its value
  in} $\alpha_{2m}=8B_0 (b_{2m,2}\!-\! L_5 c_{2m,0})/f^2$ and $\beta_{2m}=32 B_0
(b_{2m,1}\!-\! L_4 c_{2m,0})/f^2$. By isospin and charge conjugation the even
moments of different pion states (or kaon states) are equal.  Eq.~(\ref{evenMM})
implies a Gell-Mann-Okubo-like relation
\begin{equation} \label{even2MM}
 \big\langle z^{2m}\big\rangle_{\pi }+3\big\langle z^{2m}\big\rangle_{\eta}
  =4 \big\langle z^{2m}\big\rangle_{K}\ .
\end{equation}
The moment relations in Eqs.~(\ref{oddMM},\ref{even2MM}) imply relations among
the meson light cone wave functions, namely
\begin{eqnarray}
  \phi_{\pi }(x,\mu )+3\phi _{\eta }(x,\mu )
    \!\! &=&\!\! 2\big[\phi_{K^{+}}(x,\mu )+\phi_{K^{-}}(x,\mu )\big]\nn\\
 &&\hspace{-2cm}= 2\big[\phi_{K^{0}}(x,\mu )+\phi_{\overline K^{0}}(x,\mu )\big]
  \,,  \label{wa}
\end{eqnarray}
which is valid including the leading SU(3) violation.  \OMIT{where the odd
  moments cancel in the sum of kaon distributions.}  They also imply useful
relations among the frequently used Gegenbauer moments, defined by
\begin{equation}
  a_{n}^{M}(\mu )
   = \frac{4n+6}{6\!+\!9n\!+\!3n^2} 
   \int_{0}^{1}\!\!\!\!dx\,C_{n}^{3/2}(2x\!-\!1)\phi _{M}(x,\mu)\ ,
\end{equation}
with $a_0=1$.  Here $C_{n}^{3/2}(z)$ denote the Gegenbauer polynomials which are
even (odd) functions of $z$ when $n$ is even (odd). Eqs.~(\ref{oddMM}) and
(\ref{even2MM}) imply that
\begin{eqnarray}
  4a_{2m}^{K} &=& a_{2m}^{\pi }+3a_{2m}^{\eta }\ ,\qquad
  a_{2m+1}^{\pi } = a_{2m+1}^{\eta }=0 \,,\nn \\
  a_{2m+1}^{K^{0}} &=&a_{2m+1}^{K^{+}}=-a_{2m+1}^{\overline{K^{0}} }
    =-a_{2m+1}^{K^{-}}\,.  
\end{eqnarray}
In QCD or SCET factorization theorems it is often the inverse moments with respect to
the quark or antiquark that appear, $\langle x^{-1}_{q,\bar q} \rangle_M = 3 [1+
\sum_{n=1}^\infty(\pm 1)^n a_n^M]$. We find
\begin{eqnarray} \label{imom}
 &&\hspace{-0.9cm}
  \left\langle x_q^{-1}\right\rangle _{\pi }
  +3\left\langle x_q^{-1}\right\rangle_{\eta }
  =2\big[\left\langle x_q^{-1}\right\rangle _{K^{+}}
  +\left\langle x_q^{-1}\right\rangle _{K^{-}}\big] ,
  \\
 &&\hspace{-0.9cm}
  \left\langle x_s^{-1}\right\rangle _{K^-} 
   - \left\langle x_{\bar u}^{-1}\right\rangle_{K^-}
  = \left\langle x_{\bar s}^{-1}\right\rangle _{K^{+}}
   -\left\langle x_u^{-1}\right\rangle _{K^{+}} , \nn
\end{eqnarray}
with identical expressions for the antiquark $\langle x^{-1}_{\bar q}\rangle$ in
line 1 and for $\{K^-,K^+\}\to \{\bar K^0,K^0\}$ with $u\to d$ in line 2.


\begin{figure}[t!]
\vskip -0.1cm 
\centerline{
  \mbox{\epsfxsize=7truecm \hbox{\epsfbox{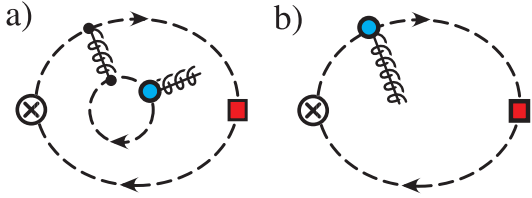}} }
  } \vskip-0.1cm
\caption[1]{Quark mass contributions to  $\phi_M(x)$ factorization. Here the
  $\otimes$ is an insertion of $O^{A,a}$, the shaded circle is ${\cal
    L}_m^{(0)}$, and the box is the meson interpolating field. Here dashed lines 
  are collinear quarks and springs are collinear gluons.}
\label{fig:scet} \vspace{-0.3cm}
\end{figure}
Finally, we relate the $m_q$ ChPT corrections to $m_q$'s in quark level
factorization in SCET. Hard-collinear factorization is simplest to derive in a
Breit frame where the quarks are collinear, with fields $\xi_n$. After a field
redefinition their LO action is~\cite{SCET}
\begin{eqnarray}  \label{L0}
  \mathcal{L}^{(0)}_{\xi\xi} = \bar \xi_n \Big[ i n \!\cdot\! D_c 
  + (i{D\!\!\!\slash}^c_\perp\!-\!m_q) \frac{1}{i {\bar n} \!\cdot\! D_c} 
  (i {D\!\!\!\slash}^c_\perp\!+\!m_q) \Big] \frac{\bar n\!\!\!\slash}{2} \xi_n 
  \,, \hspace{-0.4cm} \notag 
\end{eqnarray}
with the dependence on the matrix $m_q$ from Ref.~\cite{llw}.  The linear $m_q$
term is chiral odd and can be written
\begin{eqnarray} \label{Lm}
  {\cal L}_m^{(0)} 
    = (\bar\xi_n W) m_q\: \Big[ \frac{1}{\bnP} 
     ig {\cal B}_{\!\perp} \!\!\!\!\!\!\slash\ \Big] \: 
    \frac{\bnslash}{2}\: (W\xi_n) \,.
\end{eqnarray}
Here $ig\,{\cal B}_{\perp}\!\!\!\!\!\!\!\slash = 1/\bnP W^\dagger [ i\bn\mcdot
D_{c} , i \Dslash_{c,\perp} ] W $ and $W$ is a Wilson line of collinear
$\bn\mcdot A_n$ fields. Eq.~(\ref{Lm}) makes it clear that the Feynman rules
have $\ge 1$ collinear gluon. Similarly the chiral condensate from~\cite{llw}
can be written
\begin{eqnarray} \label{vev}
  \big\langle \Omega \big|\, (\bar\xi_{n,R}^{(i)}W) \big[ \frac{1}{\bnP} 
  ig \,{\cal B}_{\perp}\!\!\!\!\!\!\!\slash\ \: \big]\: \frac{\bnslash}{2}\: 
  (W^\dagger \xi_{n,L}^{(j)})  \big| \Omega \big\rangle 
  =  v\ \delta^{ij}\,.
\end{eqnarray}

${\cal L}_m^{(0)}$ is suppressed relative to the $m_q=0$ terms in ${\cal
  L}^{(0)}_{\xi\xi}$ by $m_q/\Lambda_{\rm QCD}$ and gives the complete set of
these corrections. Thus these corrections are universal, they
depend only on the distribution function and {\em not} on the underlying hard
process which led to $O^{A,a}$ in the first place.  (If we also want $m_q/Q$
corrections, then power corrections to $O^{A,a}$ also contribute.) Thus, at
${\cal O}(m_q/\Lambda_{\rm QCD})$ we can simply use states in the chiral limit
and add the time-ordered product
\begin{eqnarray} \label{Tsv}
 \int\!\! d^4y\:
  {\rm T}\big[ O^{A,a}(\omega_\pm)(0)\, i{\cal L}_m^{(0)}(y) \big] 
  = T_m^{(S)} + T_m^{(V)} \,.
\end{eqnarray}
The quark fields in ${\cal L}_m^{(0)}$ can either contract with themselves to
give $T_m^{(S)}$ as in Fig.~\ref{fig:scet}a (``sea'' contribution), or contract
with a quark field in $O^{A,a}$ to give $T_m^{(V)}$ as in Fig.~\ref{fig:scet}b
(``valence'' contribution). Naively the Dirac structure in Eq.~(\ref{Tsv}) cause
the matrix elements to vanish as they are odd in $\Dslash_\perp$, however the
condensate in Eq.~(\ref{vev}) makes them non-zero.  This may provide a generic
mechanism for inferring the presence of ``chirally enhanced'' condensate terms.
It also agrees with Eqs.~(\ref{oddMM},\ref{evenMM}) where $\delta\langle x^k
\rangle_M \propto \langle \bar\psi\psi \rangle$. Using Eq.~(\ref{Tsv}) and
removing $f_M$, we can work out an explicit factorization theorem where
$T_m^{(S)}$ ($T_m^{(V)}$) gives moments related to the chiral coefficient
$b_{k,1}$ ($b_{k,2}$).  This is beyond the scope of this letter.

An obvious place to study SU(3) violation in $\phi_M$ are $\pi$, $K$, and $\eta$
form factors, but also decay processes including $\bar B^0\to D_s
K^-$~\cite{mps}, $B\to M$ radiative decays, and $B\to M M'$. Our results show
that only linear $m_q$ dependence is required for lattice QCD extrapolations of
ratios of $\phi_M$ moments.  Measuring a difference between the $s$ and $u$
inverse moments for kaons in Eq.~(\ref{imom}) provides a simple way of observing
first order SU(3) violation.  The relation in Eq.~(\ref{even2MM}) could be used
to disentangle $\eta$-$\eta'$ mixing.

This work was supported in part by the Department of Energy under the
cooperative research agreement DF-FC02-94ER40818, and by the National Science
Council of ROC. I.S.  was also supported by a DOE Outstanding Junior
Investigator award. Preprint MIT-CTP 3445.

\end{document}